\documentclass[prb,twocolumn,amsmath,amsfonts,10pt]{revtex4-1}

\usepackage{epsf}

\usepackage{graphicx}

\usepackage{color}

\usepackage{amssymb}

\usepackage{amsmath}

\usepackage{revsymb4-1}

\usepackage{rotating}

\usepackage{tabulary}

\usepackage{setspace}

\usepackage{multirow}

\usepackage{color}

\usepackage{wasysym}

\definecolor{green0}{rgb}{0,0.5,0}

\usepackage{bm}

\usepackage{fixmath}

\usepackage{amsbsy}

\usepackage{pifont}

\usepackage{diagbox}

 \definecolor{darkgreen}{rgb}{0,0.5,0} % VM

 \usepackage{color}

 \usepackage[normalem]{ulem}

 \usepackage{cancel}

\newcolumntype{C}{>{\centering\let\newline\\\arraybackslash\hspace{0pt}}m{0.8cm}}

\newcolumntype{D}[1]{>{\centering\let\newline\\\arraybackslash\hspace{0pt}}m{#1}}

\newcolumntype{E}{>{\centering\let\newline\\\arraybackslash\hspace{0pt}}m{0.9cm}}

 \definecolor{darkgreen}{rgb}{0,0.5,0} % VM

 \usepackage{color}

 \usepackage[normalem]{ulem}

 \usepackage{cancel}

\begin{document}

\title{Advanced density matrix renormalization group method\\ for nuclear structure calculations}

%RevTeX form:

\author{\"O. Legeza$^{1}$, L. Veis$^{1,2}$, A. Poves$^3$, J.Dukelsky$^{4}$}

\affiliation{
   $^1$ Strongly Correlated Systems ``Lend\"ulet'' Research group,
   Wigner Research Centre for Physics, H-1525 Budapest, Hungary\\
   $^2$ J. Heyrovsk\'y Institute of Physical Chemistry, Academy of Sciences of the Czech Republic (ASCR), CZ-18223 Prague, Czech Republic\\
   $^3$ Departamento de F\'{i}sica Te\'{o}rica e IFT-UAM/CSIC, Universidad Aut\'{o}noma de Madrid, 28049 Madrid, Spain\\
   $^4$ Instituto de Estructura de la Materia, CSIC, Serrano 123, 28006 Madrid, Spain}

\date\today

%\vskip -8pt

\begin{abstract}

We present an efficient implementation of the
Density Matrix Renormalization Group (DMRG) algorithm that includes an
optimal ordering of the proton and neutron orbitals and
an efficient expansion of the active space
utilizing various concepts of quantum information theory.
We first show how this new DMRG methodology could solve a previous $400$ KeV discrepancy in the ground state energy of $^{56}$Ni.
We then report the first DMRG results in the $pf+g9/2$ shell model space for the ground $0^+$ and first $2^+$ states of  $^{64}$Ge
which are benchmarked with reference data obtained from
Monte Carlo shell model. The corresponding correlation structure among the proton
and neutron orbitals is determined in terms of the two-orbital mutual
information. Based on such correlation graphs we propose several further
algorithmic improvement possibilities that can be utilized in a new generation of tensor network
based algorithms.
\end{abstract}


\maketitle

Large scale Shell Model calculations in physically sound valence spaces are the prime choice
in the nuclear spectroscopy of light and medium mass nuclei. The catch is that the dimensions of the basis
(the number of Slater determinants) grow as the product of the two combinatorial numbers made with the number
of single particle states ($n,l,j,m$) in the valence space and the number of particles, for neutrons and protons.
Modern shell model codes can cope with dimensions ${\mathcal O}(10^{11}$) \cite{Caurier2005}. Going beyond this limit requires clever truncation and extrapolation schemes. In addition,  different approximation methods may be implemented.
Among them, perhaps the most accurate is the Monte Carlo Shell Model (MCSM) that stochastically samples the Hilbert space to find the relevant basis states for the description of a particular eigenstate \cite{Otsuka1995}. Several improvements have been added to the original methodology along the years making the MCSM a robust and accurate technique to study medium-mass nuclei. The most important are sequential conjugate gradient to select the Slater determinants and the parity and angular momentum projection of the determinants, plus an energy variance extrapolation method \cite{Otsuka2010}.

Another efficient numerical tool to approximate the exact wave function
in a truncated basis, is the DMRG
method \cite{White-1992} that was earlier introduced in nuclear structure either in the particle-hole basis (phDMRG) \cite{Duk2002, Duk2004} or in the j-coupling scheme (JDMRG) \cite{Duk2004, Pit2006}. Both methods found difficulties to treat systems beyond the limits of an exact diagonalization. However, phDMRG in the j-coupling scheme has been successfully applied to the Gamow shell model of weakly bound light \cite{Duk2006, Rot2009} nuclei due to the weak entanglement between the valence space and resonances with the discretized continuum. Moreover, a standard implementation of DMRG to nuclear structure found a serious  discrepancy of $~400$ KeV for the $^{56}$Ni ground state  energy in the $pf$ shell \cite{Pape2005}. These results hindered further application of the DMRG method to medium size nuclei.
Meanwhile, there have been important efforts to implement DMRG in quantum chemistry (QC-DMRG) \cite{White-1999} by utilizing various concepts
of quantum information theory
\cite{Legeza-2003c,Rissler-2006,Barcza-2011,Kurashige-2013}.
Nowadays, DMRG is capable to provide the low-lying energy spectrum of complex molecules with great
accuracy \cite{Legeza-2008,Reiher-2010,Chan-2011,Wouters-2014,Szalay-2015}
and it has ranked among the standard multireference QC methods.

In this letter we will make use of the new DMRG techniques developed in QC to overcome first the $400$ KeV discrepancy found in the $^{56}$Ni calculations  in the $pf$ shell. Afterwards, we will show how DMRG can deal with nuclei well beyond the limits of an exact diagonalization by studying
 $^{64}$Ge in the $pf$ and in the enlarged $pf+g9/2$ valence space, and compare our results with benchmark calculations of MCSM. As a byproduct of the method, we will depict the single-site entropy and mutual information of  $^{64}$Ge, shedding new light into the landscape of entanglement
 and correlations in nuclear structure.

In our DMRG implementation we study the most general Hamiltonian with one- and
two-body interaction terms given as
\begin{equation}
\label{eq:ham}
H = \sum_{\alpha} \varepsilon_{\alpha} c_{\alpha}^{\dagger} c_{\alpha} -
\frac{1}{2} \sum_{\alpha \beta \gamma \delta } V_{\alpha \beta \gamma \delta} c_{\alpha}^{\dagger} c_{\beta}^{\dagger} c_{\delta} c_{\gamma}\,,
\end{equation}
where $c_{\alpha}^{\dagger}$ and $c_{\alpha}$ creates and annihilates
a particle with quantum numbers $\alpha=(n,l,j,m,\tau_z)$.

In the so-called ${\mathbb C}^2$ representation an
orbital can be either empty or occupied, thus
the dimension of the local Hilbert space, $\Lambda_i$ of a single orbital,
$q={\rm dim}\ \Lambda_i$ is 2.
The full Hilbert space of a finite system comprising $N$
orbitals, $\Lambda^{(N)}$,
is built from tensor product spaces of local orbital spaces
$\Lambda_i$, which can be written as
$\Lambda^{(N)}=\otimes_{i=1}^N \Lambda_i$.

In the DMRG method, quantum correlations are taken into account by an iterative
procedure that  variationally minimizes the energy of
the Hamiltonian given by Eq.~(\ref{eq:ham}). The method eventually
converges to the full Configuration Interaction (CI)   solution within the selected active orbital
space.
In the two-site DMRG variant \cite{White-1992,Schollwock-2005},
$\Lambda^{(N)}$  is approximated by a tensor product
space of four tensor spaces defined on an ordered orbital chain, i.e.,
$\Xi^{(N)}_{\rm DMRG}=\Xi^{(\mathrm{l})}\otimes\Lambda_{i+1}\otimes\Lambda_{i+2}\otimes\Xi^{(\mathrm{r})}$.
The basis states of the $\Xi^{(\mathrm{l})}$
comprises $i$ orbitals to the left of the chain ($l \equiv$ left ) and  $\Xi^{(\mathrm{r})}$ comprises $N-i-2$ orbitals
to the right of the chain ($r \equiv$ right). These states
are determined  through a series of unitary transformation based on
the singular value decomposition (SVD) theorem
by going through the ordered orbital space from {\it left} to {\it right} and then sweeping back and forth
 \cite{Schollwock-2005,Szalay-2015}.
The number of block states, $M_l={\rm \dim}\ \Xi^{(\mathrm{l})}$
and $M_r={\rm \dim}\ \Xi^{(\mathrm{r})}$, required to achieve
sufficient convergence can be regarded as a function of the level of
entanglement among the orbitals \cite{Vidal-2003,Legeza-2003c}.
The maximum number of block states
$M_{\rm max} = \max{(M_l,M_r)}$\ required to reach an
\emph{a priory} defined  accuracy threshold,  is inherently determined by
truncation error, $\delta \varepsilon_{\rm TR}$,
when the Dynamic Block State Selection (DBSS) approach is
used \cite{Legeza-2003a}.
During the initial sweeps of the DMRG  algorithm  the
accuracy is also affected by the
environmental error, $\delta \varepsilon_{\rm sweep}$.
The latter error can be reduced significantly by taking advantage of
the CI based  Dynamically  Extended Active Space
procedure (CI-DEAS) \cite{Legeza-2003c,legeza2004-leiden}
and using a large number of DMRG sweeps
until the energy change between
two sweeps is negligible. 
In the CI-DEAS procedure the active space of orbitals is extended
dynamically based on the orbital entropy
profile \cite{Barcza-2011}.
$M_{\rm max}$
depends strongly on the orbital ordering along the
one-dimensional chain topology of the DMRG
method \cite{Legeza-2003a,Moritz-2005}.
There  exist various extrapolation schemes to determine the
truncation-free solution \cite{Wouters-2014}.
In  this work, due to the high level of entanglement of the nuclear wave functions,
we carry out the extrapolations
as a function of the total number of block states $M$.
Using $E(M)=E(M\rightarrow\infty)+X_1M^{X_2}$ we have estimated the
truncation-free solution where
$E(M\rightarrow\infty), X_1$ and $X_2$ are free parameters of our fit~\cite{Barcza-2013}.
We have performed between 30 to 90 sweeps
by requiring the energy change between two sweeps to be below $10^{-4}$
MeV.

The amount of contribution to the total correlation energy of an orbital
can be quantified  by the single-orbital von Neumann entropy\cite{Legeza-2003c},
$s_i=-{\rm Tr} \rho_i \ln \rho_i$ where $\rho_i$ is the reduced density
matrix of  orbital $i$. The two-orbital von Neumann entropy $s_{ij}$ is constructed
similarly using the reduced density matrix, $\rho_{ij}$ of a subsystem built
from orbitals $i$ and $j$,  and the mutual information
$I_{ij}=s_{ij}-s_i-s_j$ describes how orbitals are correlated
with each other as they are embedded in the whole system.~\cite{Rissler-2006,Barcza-2011,Boguslawski-2013a}
The orbital ordering is determined by the minimization of the
entanglement distance expressed as $I_{\rm dist} = I_{ij} \: |i-j|^\eta$
where $\eta\ge1$.~\cite{Rissler-2006,Barcza-2011}. 
In this work we used $\eta=2$ in order to carry out the optimization task
using concepts of spectral graph theory~\cite{atkin98}. 

\begin{figure}[!htb]
\centerline{
 \includegraphics[scale=0.6]{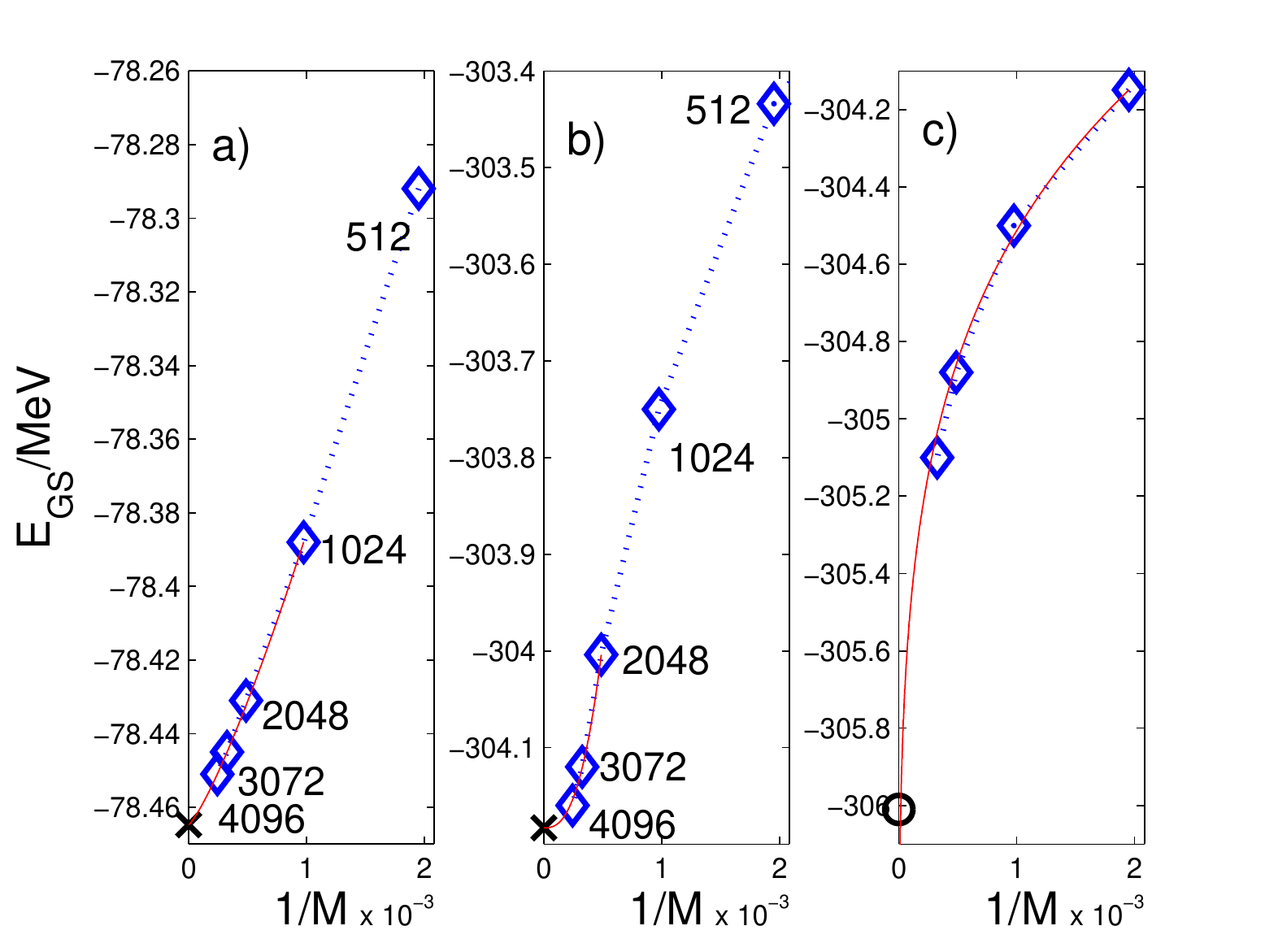}}
\caption{(Color online)
The DMRG ground state energy for $^{56}$Ni and
$^{64}$Ge in  the $pf$-shell  and $^{64}$Ge in the $pf+g9/2$ valence space, are shown as a function of
$1/M$. Diamonds (DMRG), Crosses (Exact Diagonalization), Circle (MCSM).
The solid lines are the fits that produce our extrapolated value, while the dotted lines
 joining the points not used in the fit  are just to guide the eye.
}
\label{fig:energy}
\end{figure}

Our DMRG code for nuclear structure calculations is composed of two phases.
The {\em preprocessing phase}
in which the orbital ordering and  the active space vectors are optimized
by calculating the one-orbital entropy and two-orbital mutual information
using a fixed small number
of block states.  The {\em production phase} in which an
accurate calculation is performed using large fixed number of block states or the DBSS procedure
in order to reach an \textit{a priory} set  accuracy threshold ~\cite{Fertitta-2014}.
The preprocessing phase takes only a small fraction of the total computational
time.

First we perform the  DMRG calculation for the $sd$ nucleus
$^{28}$Si  employing  the USD interaction \cite{ref_usd_19}.
This system was treated almost a decade ago by the DMRG method \cite{Pape2005}.
By keeping fixed $M=1024$ block states
we have reproduced the exact value of the 0$^+$ ground state energy
and excitation energy of the first $2^{+}$  state.
Using the quantum information entropy based ordering optimization,
the CI-DEAS procedure, and the DBSS
approach with $M_{\rm  min}=256$ and 
$\delta \varepsilon_{\rm TR}=10^{-3}-10^{-6}$
we determined the ground state energy  within the \textit{a priory set}  
accuracy threshold.
This drastic improvement
clearly  demonstrates the importance of the quantum entropy based optimization
procedures.

Next we consider the nucleus $^{56}$Ni with the KB3
interaction \cite{ref_kb3_20,ref_kb3_21}.
The exact results  in $pf$ valence space  were obtained by the code ANTOINE \cite{Caurier2005}.
The basis dimension  in m-scheme for $J_z$=0 is
$1.1\times 10^9$.  
In Fig.~\ref{fig:energy}(a) the DMRG ground state energy for $^{56}$Ni
is shown as a function of $1/M$.
For small $M$ values a downward
curvature governs the behavior of the scaling function
(indicated by the dotted lines) while for large
enough $M$ values,  an inflection point is reached and an upward curvature
becomes apparent.
Our lowest variational energy obtained with $M=4096$ states is
$E_{\rm GS}^{(\rm Ni)}(M=4096)=-78.451$~MeV, thus the error compared to the
exact value, $E_{\rm exact}=-78.465$~MeV, is $1.4\times 10^{-2}$~MeV.

Using the fit function defined above we have estimated the
ground state energy in the $M\rightarrow\infty$ limit
as  $E_{\rm GS}^{(\rm Ni)}(M\rightarrow\infty)=-78.463(4)$~MeV.
The deviation from the exact ground state energy
is in the order of a few keV's.
Therefore, using entropy based optimized DMRG methodology we could solve
the previous $400$~keV discrepancy in the ground state energy of $^{56}$Ni
reported in Ref.~[\onlinecite{Pape2005}].

Next we study $^{64}$Ge in the $pf$ valence space for
which we have used the GXPF1A interaction \cite{Honma2004, Horoi}.
The DMRG results for the $pf$ valence space are
displayed in Fig.~\ref{fig:energy}(b).
Our lowest variational energy obtained with $M=4096$ states is
$E_{\rm GS}^{(\rm Ge)}(M=4096)=-304.163$~MeV  which  extrapolates to
$E_{\rm GS}^{(\rm Ge)}(M\rightarrow\infty)=-304.182(3)$~MeV
while the exact energy obtained with the code ANTOINE is $E_{\rm exact}=-304.183$~MeV. Therefore, the error
is again in the keV range.

Being confident with the results obtained in the $pf$ shell, we proceed to explore the ability of DMRG to describe accurately the structure of nuclei in valence spaces that exceed the limits of an exact diagonalization. With this in mind, we study $^{64}$Ge in the extended space $pf+g9/2$
(dimension in m-scheme $1.7 \times 10^{14}$)  which was already considered using  the MCSM method \cite{Otsuka2010, Otsuka2012}. In this case the GXPF1A interaction has to be completed with the matrix elements
between the $pf$-shell  orbits and the $g9/2$  obtained in a standard G-matrix calculation.
In addition,  the spurious center of mass  contamination is treated with the Lawson's  ansatz  \cite{Horoi,Honma}.
 Within this extended valence space we found a significantly
slower convergence rate as a function of $M$ as can be seen in
Fig.~\ref{fig:energy}(c). Even using up to $M=3072$
block states the inflection  could not be reached, thus the upward curvature
did not become apparent yet. Therefore, the estimated energy
using our fit function overshoots the reference MCSM energy  $(E_{\rm MCSM}=-306.066)$~MeV
and can provide only
a lower bound which we found to be
$E_{\rm GS}^{(\rm Ge)}(M\rightarrow\infty)=-307.4$~MeV. An upper bound can be estimated
using a second order polynomial fit giving
$E_{\rm GS}^{(\rm Ge)}(M\rightarrow\infty)=-305.5$~MeV. The extrapolated
MCSM reference energy lies within the two bounds
given above.
Due to the slow scaling of the energy as a function
of $1/M$  significantly more block states are needed
to provide a reliable extrapolation.

\begin{figure}[!htb]
\centerline{
  \includegraphics[scale=0.78]{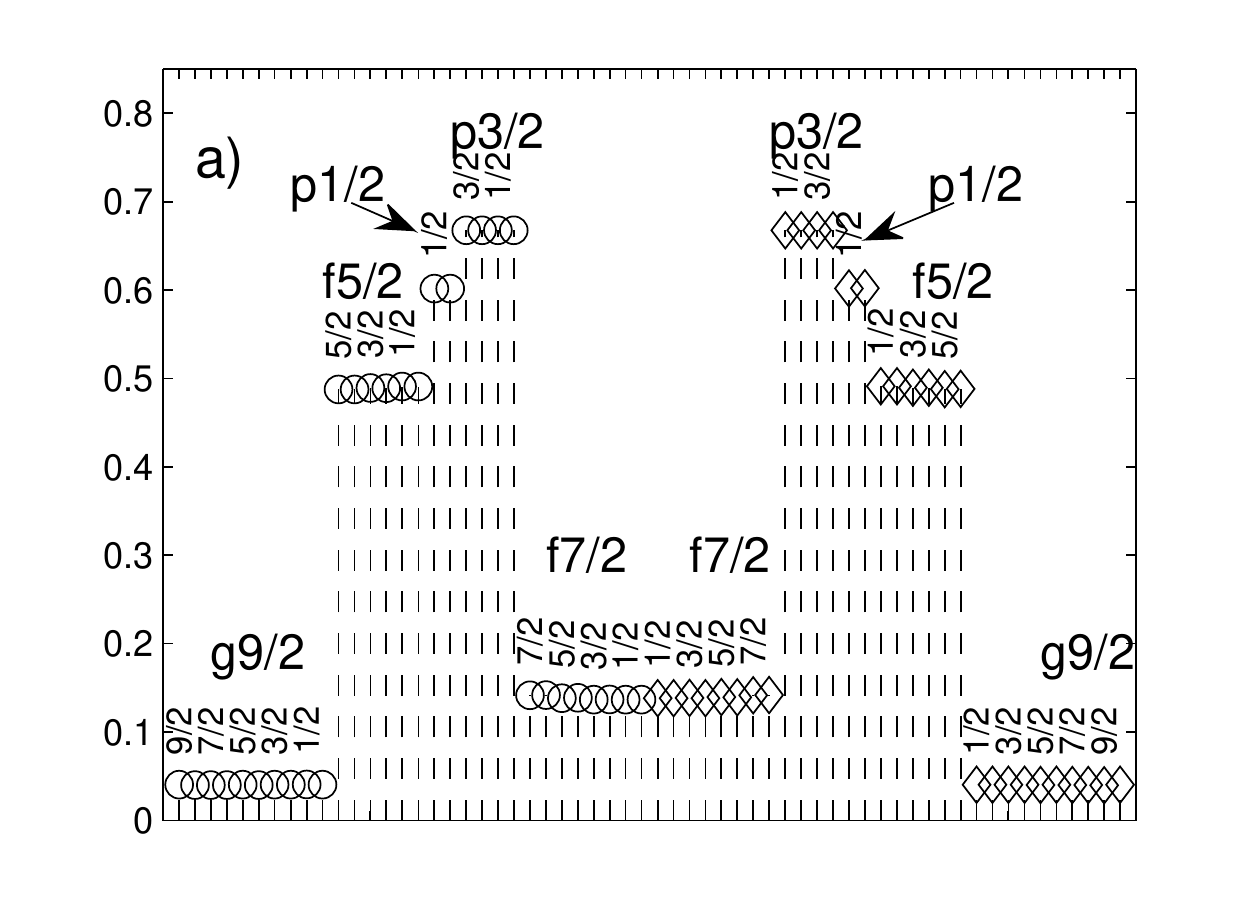}
}
\vskip -0.5cm
\centerline{
  \includegraphics[scale=0.78]{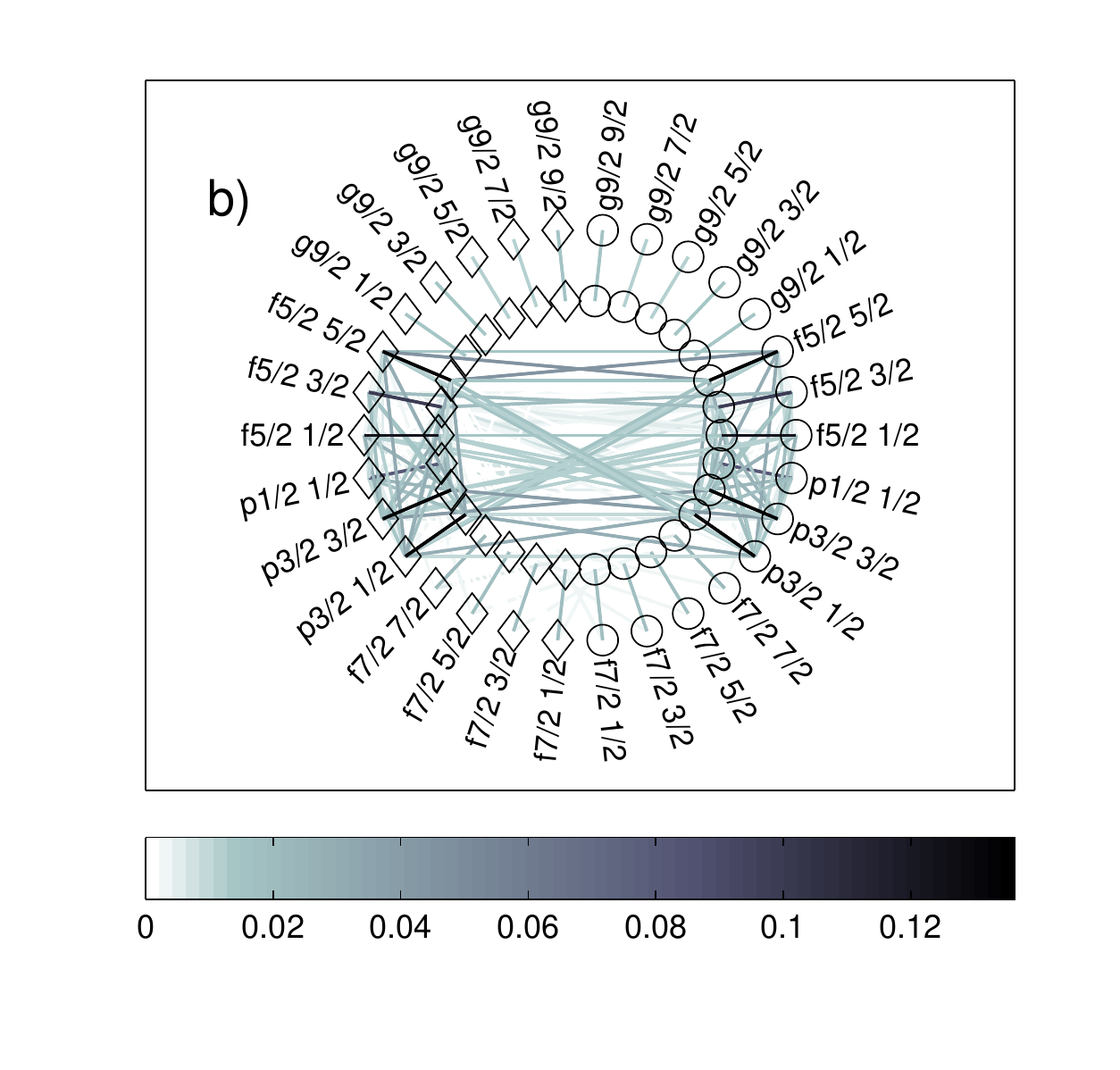}
}
\caption{(Color online)
Single-site entropy and mutual information obtained with  DMRG
for $^{64}$Ge in the $pf+g9/2$-shell. The mutual information matrix
elements are shown on a ladder topology
with time-reversed pairs in the rungs. Circles and diamonds label
proton and neutron orbitals respectively, and
sites are denoted by  $l,j,m$ with $+m$ outside the ladder and $-m$ inside.
}
\label{fig:mutual_Ni}
\end{figure}

Besides optimization procedures the entanglement analysis is also very
important to obtain physical information encoded in the wave functions
\cite{Boguslawski-2012b,Boguslawski-2013a,Kurashige-2013,Fertitta-2014}.
The single-orbital entropy and two-orbital mutual information obtained
with  DMRG for  $^{64}$Ge
are shown in Fig.~\ref{fig:mutual_Ni}.

The single orbital entropy profiles reflect the strong configuration mixing
in the wave function. The magnetic states  belonging to the same orbital
have the same entropy  (within the numerical accuracy). The entropies
of the  $p1/2$ and $p3/2$ orbitals are very close to the upper limit $\ln 2=0.693$
which corresponds to the maximally mixed state. Orbitals
with largest entropies contribute the most to the correlation energy
thus their accurate treatment is mandatory. As expected, the entropy profiles show that the $p1/2$, $p3/2$ and $f5/2$ orbits play a dominant role in the physics of the system, whereas the presence of the $f7/2$ orbit is non-negligible and that of the $g9/2$ is minor. Notice that these three highly entangles orbits close a pseudo-SU(3) symmetry which enhances the quadrupole-quadrupole correlations as it will be apparent in the mutual information diagram.

The two-orbital mutual
information shows how orbitals are correlated with each other.
The rungs of the ladder display the degree of entanglement between time reversed states of like particles. The fact that the mutual information is approximately equal for the $p1/2$, $p3/2$ and $f5/2$ orbits, and independent of their $j_z$ projections, is consistent with the presence of a strong T=1 proton-proton and neutron-neutron pairing coherence. The same coherence, though significantly less intense, is seen in the $f7/2$ and the $g9/2$. Proton-neutron T=1 pairing correlations could also be seen between time reversed and charge conjugated states for the $p1/2$, $p3/2$ and $f5/2$ orbits. We could also see significant entanglement between proton-neutron maximally aligned states for the $p3/2$ and $f5/2$ orbits, which could be related to $J=2j_z$ pairing and/or quadrupole-quadrupole in the T=0 channel. Similarly, quadrupole correlations can be observed inside the ladder (proton-proton and neutron-neutron) as well as those connecting opposite sites (proton-neutron) of the ladder for the  $p1/2$, $p3/2$ and $f5/2$ orbits.

Let's consider now the excitation energies of the $2^+$ states.
It is worth to remark that even in the case when the ground state
energy could not be determined with the desired accuracy, the energy difference
between two states can be obtained more accurately due to cancelation of errors by targeting
the two lowest lying eigenstates simultaneously.
The exact value of the  $2^+$  excitation energy for $^{28}$Si can be reproduced
with  $M=1024$ block states.
For $^{56}$Ni, using
fixed $M=2048$ block states,  we obtain  $\Delta E=5.218$~MeV to be compared with the exact value
 $5.125${ MeV,  with an  absolute error  of  93~keV.
For $^{64}$Ge in the $pf$-shell  we get  $\Delta E=0.922$~MeV  compared with the exact value $0.906$~MeV, the absolute error
being  now 16~keV. Using the fitting procedure to extrapolate  the $2^+$  excitation energies to M=$\infty$
we obtain
$5.121(3)$~MeV  and $0.907(2)$~MeV for  $^{56}$Ni and  $^{64}$Ge.  Indeed the agreement is excellent.

When the $pf+g9/2$ valence space is considered we found a slower
convergence rate
as a function of $1/M$ due to the more complex entanglement structure in
the system. Fig.~\ref{fig:gap}(a) and (b) show the $2^+$ excitation energies of  $^{64}$Ge
as a function of $1/M$ for the $pf$-shell and
$pf+g9/2$-shell, respectively. The solid lines represent the fitted values.
In the latter case the exact solution is not available, thus
the extrapolated MCSM energies
\cite{Otsuka2012},
$\Delta E=0.919$ (without reordering) or
$\Delta E=0.890$ (with reordering)
can be taken as a benchmark reference and are  indicated by the blue diamond symbol.
Our extrapolated excitation energy
$\Delta E=0.90(2)$ is in good agreement with the MCSM result.

\begin{figure}[!htb]
\centerline{
 \includegraphics[scale=0.6]{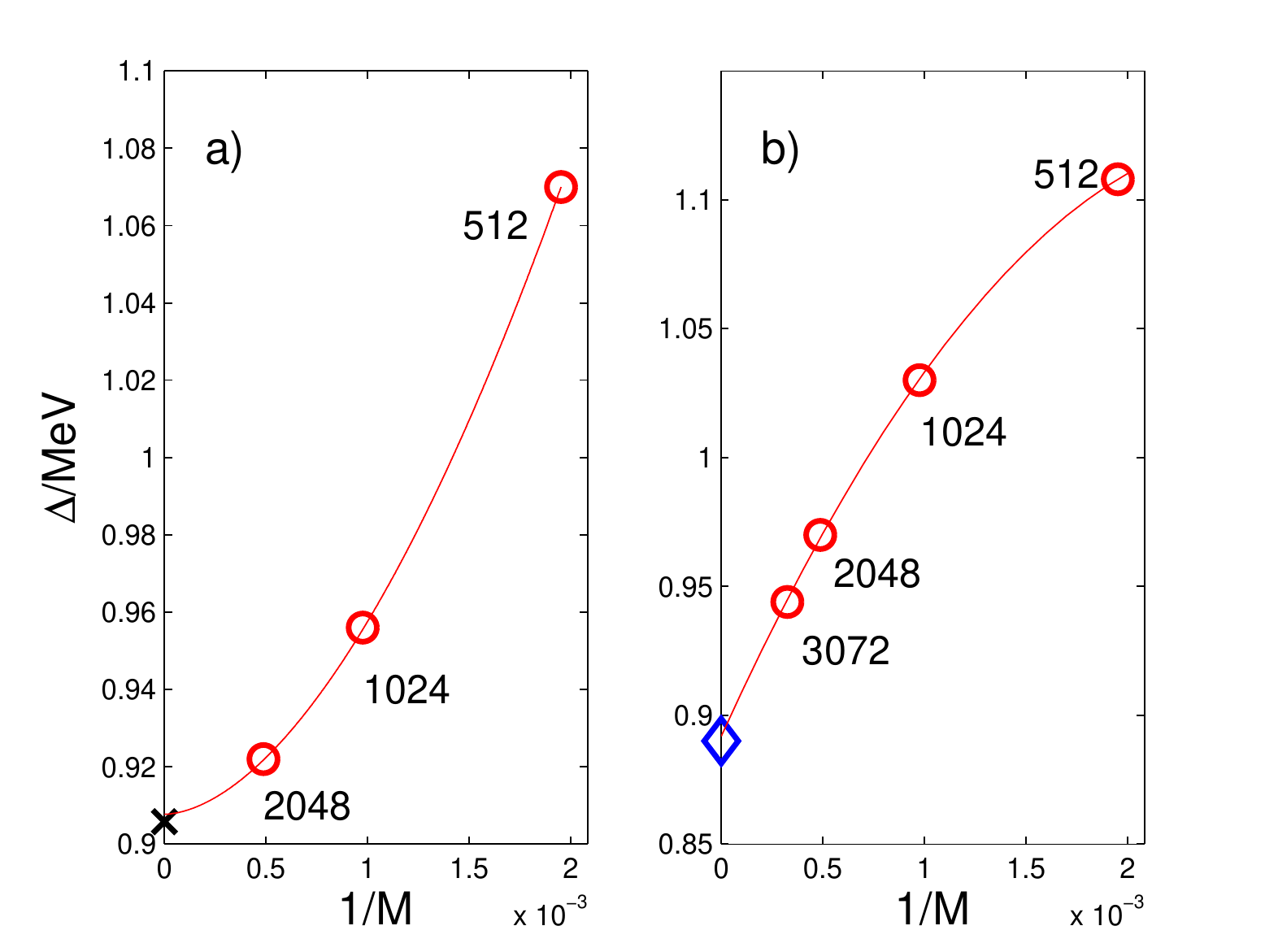}
}
\caption{(Color online)
The energy gap between the $0$ and $2^{+}$ states for $^{64}$Ge with
$pf$ and $pf+g9/2$ valence space
are shown as a function of $1/M$.  
Circles (DMRG), 
Cross (Exact Diagonalization), 
Diamond (MCSM). The
solid lines are our fits. 
}
\label{fig:gap}
\end{figure}

In this work we have demonstrated that the DMRG method including
various novel optimization algorithms based on quantum
information theory can be applied efficiently to nuclear structure calculations for medium-mass nuclei in extended valence spaces with dimensions exceeding the limits of an exact diagonalization.
The DMRG results for the ground state energy
 of $^{64}$Ge in the $pf+g9/2$ valence space
could be  further improved in order to obtain a reliable
estimate for the truncation-free limit. However, the excitation energy of the first 2$^+$ is in excellent agreement
with MCSM benchmark results. By calculating the single-orbital and
two-orbital mutual information we have determined the correlation
structure among the orbitals. These  entanglement graphs, based on recent developments in QC  applications \cite{Szalay-2015}, constitute a novel and highly precise tool to picture the wave function correlation properties. The analysis of these entanglement graphs suggests new alternatives for efficient truncation methods that can be developed on general grounds. A straightforward extension of the DMRG algorithm could be the use of local tensors with $\Lambda_i$ of dimension $q=4$, but unlike the QC implementations, the states to include in this tensor should be time-reversed pairs in order to optimize the treatment of nuclear pairing correlations.
As a final remark, the use of tree-tensor network state (TTNS), which is a recent development in quantum information theory \cite{Murg-2010-tree,Nakatani-2013-tree,Murg-2015-tree} could tackle the problem of having equal single entropy values for each orbital group that makes inefficient a sequential treatment of these states within an ordered chain.  Within this scheme, several orbitals with equally large entropy values could form the central shells of the TTNS network.
These new developments could certainly open the field of medium to heavy nuclei to highly precise spectroscopic calculations.

\begin{acknowledgements}

We thank the hospitality of NORDITA during the workshop "Computational Challenges in Nuclear and Many-Body Physics" when this project started. We acknowledge the invaluable help of M. Horoi and T. Papenbrock in the initial part of the project.
This work was supported in part by the
Hungarian Research Fund (OTKA) through Grant Nos.~K100908 and ~NN110360, and by the Spanish Ministry of Economy and Competitiveness through Grants FIS2012-34479, FPA2011-29854 and SEV2012-0249.

\end{acknowledgements}

\end{document}